\documentclass[a4paper,12pt]{article}
\usepackage{epsfig}
\usepackage[dvips,usenames]{color}
\usepackage{graphicx}

\newlength{\dinwidth}
\newlength{\dinmargin}
\newcommand{\spur}[1]{\not\! #1 \,}
\begin{document}
\title{Revisiting the annihilation  decay
 $\bar{B}_{s}\to\pi^{+}\pi^{-}$ }

\bigskip
\author{Yadong Yang\footnote{ Corresponding
author. E-mail address: yangyd@henannu.edu.cn},~ Fang Su,
 Gongru Lu and Hongjun Hao
\\
{ \small \it Department of Physics, Henan Normal University,
Xinxiang, Henan 453007,  P.R. China}\\
}
\bigskip\bigskip
\maketitle

\begin{abstract}
\noindent It is very important to know the strength of
annihilation contribution in B charmless nonleptonic decays.
$\bar{B}_{s}\to \pi^{+}\pi^{-}$ process could serve a good probe
of the strength. We have studied the process in QCD factorization
framework. Using a gluon mass scale indicted by the studies of
infrared behavior of gluon propagators to avoid enhancements in
the soft end point regions, we find that the CP averaged branching
ratio is about $1.24\times 10^{-7}$, the direct CP asymmetry
$C_{\pi\pi}$ is about -0.05, while the mixing-induced CP asymmetry
quite large with the value $S_{\pi\pi}$=0.18. The process could be
measured at LHC-b experiments in the near future and would deepen
our understanding of dynamics of B charmless decays.
\end{abstract}
\bigskip\bigskip

\newpage
\section{Introduction}
\noindent Recent years many efforts have been made to understand
charmless decays of B mesons, which provide  good grounds to get
deep insights into the flavor structure of the Standard Model
(SM), the origin of CP violation, the dynamics of hadronic decays,
and to search for any signals of new physics beyond the SM. Up to
now, BaBar(SLAC)~\cite{babar} and Belle(KEK)~\cite{belle} have
already accumulated large set of data and have made  plenty of
exciting measurements. Moreover, in the near future LHC-b
experiment, the expected number of $b\overline{b}$ events produced
per year is about $10^{12}$, it is  noted that 10\% of the events
would fragment to $B_{s}$ mesons. This high statistics will allow
studies of rare $B_s$ decay modes, which will  provide very
sensitive tests of theories for B decays, electro-weak interaction
models and so on.

For nonleptonic B meson decays, the most difficult aspect lies in
the computation of matrix elements of the effective four-quark
operators between hadron states. To deal with this, a simple and
widely used approach is the so-called factorization
approach(FA)~\cite{Wirbel}. In the past few years, new approaches,
such as the QCD factorization(QCDF)\cite{M} and perturbation QCD
(pQCD) scheme\cite{lihn} have been proposed to improve the FA on
QCD grounds.

In the most cases of B meson nonleptonic decays, the annihilation
contribution carries  weak and strong phases different from that
provided by the tree or penguin amplitudes, which is very
important for studying  CP-violating observables. Meanwhile, the
calculation of annihilation contributions is interesting by
itself, since it can help us to understand the low energy QCD
dynamics and the viability of the theoretical approaches. As
argued in~\cite{M}, the annihilation amplitude is formally power
suppressed by order $\Lambda_{QCD}/m_{b}$ in QCDF.  However, the
annihilation contribution may not be small. In a recent systematic
calculation of B decays\cite{bn}, it is shown that the
annihilation contributions could cause considerable uncertainties
in their theoretical predictions, where the contributions are
parameterized in term of the divergent integral $\int^1_0
\frac{dy}{y}\to X_{A}=(1+\varrho_A
e^{i\varphi})\ln\frac{m_B}{\Lambda_h} $. In this paper, we argue
that the strength of annihilation could be probed by measuring the
interesting decay mode $\bar{B}_s \to \pi^+\pi^-$, which is a pure
annihilation process. In our calculation of the scattering kernel,
we will use Cornwall\cite{Cornwall} prescription of gluon
propagator with a dynamical mass to avoid enhancements in the soft
end point. It is very interesting to note that recent
theoretical\cite{thstudies} and phenomenological\cite{phstudies}
studies are now accumulating supports for softer infrared behavior
for gluon propagator. Besides serving a probe for the
annihilation, the decay has some interesting features: sizable CP
violation due to both tree and penguin operators contributing,
clear experimental signatures due to its two charge final states.
Moreover, if few percentage of final pions are mis-identified to
be muons, it would bring considerable uncertainties  to the
measurement of $\bar{B}_s \to \mu^+ \mu^-$ at LHCb. Therefor, the
decay deserves our theoretical studies using different approaches.

 We have found that the CP averaged branching ratio of
$\overline{B}_{s}\to\pi^{+}\pi^{-}$ decay is about $1.24\times
10^{-7}$, the direct CP asymmetry $C_{\pi\pi}$ is about -0.05,
while the mixing-induced CP asymmetry is as large as
$S_{\pi\pi}$=0.18. Our results might be tested in the near future
at LHCb.

The remaining parts of this paper are organized as follows. In the
next section, we outline the necessary ingredients of the QCD
factorization approach for describing the
$\overline{B}_{s}\rightarrow\pi^{+}\pi^{-}$ decay and calculate
the decay amplitude. In section 3, we give the numerical results
of the CP averaged branching ratio  and discuss CP asymmetries in
$\overline{B}_{s}\rightarrow\pi^{+}\pi^{-}$ decay.

\section{$\bar{B}_{s}\to\pi^{+}\pi^{-}$ decay in QCD factorization approach}
We will start as usual from the  effective Hamiltonian for the
$\bigtriangleup B=1$ transitions given by~\cite{Buchalla}
\begin{eqnarray}
{\cal
H}_{eff}&=&\frac{G_{F}}{\sqrt{2}}\{V_{ub}V_{us}^*[C_{1}(\mu)O_{1}(\mu)+
C_{2}(\mu)O_{2}(\mu)]
-V_{tb}V_{ts}^*\sum\limits_{i=3}^{10}C_{i}(\mu)O_{i}(\mu)\}+h.c,
\end{eqnarray}
where $C_{i}$ are Wilson coefficients at the renormalization scale
$\mu$ in the Standard Model by integrating out heavy gauge bosons
and top quark fields. $O_{1,2}$ are  tree operations arising from
W-boson exchange and $O_{3-10}$ are penguin operators.  The values
for $C_i$  and the definition of operators $O_{i}$ could be found
in \cite{Buchalla}.

With the effective Hamiltonian,  the amplitude for
$\bar{B}_{s}\to\pi^{+}\pi^{-}$ in naive factorization is
\begin{eqnarray}
{\cal
A}(\overline{B}_{s}\to\pi^{+}\pi^{-})&=&-~2\frac{G_{F}}{\sqrt{2}}V_{tb}V_{ts}^*
\biggl[(a_3+\frac{3}{2}Q_u
 a_9)\langle \pi^{+}\pi^{-}|\overline{u}\gamma_{\mu}Lu|0\rangle
\langle 0|\overline{s}\gamma^{\mu}Rb|\overline{B}_{s}\rangle\nonumber \\
&~~~~~&+~(a_{5}+\frac{3}{2}Q_{u}a_7) \langle
\pi^{+}\pi^{-}|\overline{u}\gamma_{\mu}Ru|0\rangle
\langle 0|\overline{s}\gamma^{\mu}Lb|\overline{B}_{s} \rangle \biggl]
\nonumber\\
&~~~~~&+~\frac{G_{F}}{\sqrt{2}}V_{ub}V_{us}^* a_2 \langle
\pi^{+}\pi^{-}|\overline{u}\gamma_{\mu}Lu|0\rangle
\langle 0|\overline{s}\gamma^{\mu}Lb|\overline{B}_{s}\rangle+(u\to d)\nonumber\\
&=&-~2i\frac{G_{F}}{\sqrt{2}}V_{tb}V_{ts}^*f_{B_s}p_B^\mu
\biggl[(a_3+\frac{3}{2}Q_u a_9)
\langle\pi^{+}\pi^{-}|\overline{u}\gamma_{\mu}Lu|0\rangle\nonumber\\
&~~~~~&+~(a_{5}+\frac{3}{2}Q_u a_7)\langle\pi^{+}\pi^{-}|\overline{u}
\gamma_{\mu}Ru|0\rangle\biggl]\nonumber\\
&~~~~~&+~i\frac{G_{F}}{\sqrt{2}}V_{ub}V_{us}^*f_{B_s}p_B^\mu a_2
\langle \pi^{+}\pi^{-}|\overline{u}\gamma_{\mu}Lu|0\rangle+(u\to d
),
\end{eqnarray}
where $L,R=(1\mp\gamma_5 )/2$.  Due to the conservation of vector
current and partial conservation of axial-vector current, this
amplitude will vanish in the limit $m_{u},m_{d}\rightarrow 0$. To
$\alpha_{s} $ order, the  matrix $\langle
\pi^{+}\pi^{-}|u\spur{P}_{B}(1-\gamma_{5})u|0\rangle$ also
vanishes due to the cancellation between the amplitudes of Fig.1
(a) and (b). So that, nonfactorizable contribution will  dominate
the decay, which can be obtained by calculating the amplitudes of
Fig.1 (c) and (d). We consider the contribution up to  the twist-3
distribution amplitude of the light mesons which is superficially
suppressed by $\mu_{\pi}$, however, $\mu_{\pi}$ is much larger
than its naive scaling estimation $\Lambda_{QCD}$ \cite{M}
\begin{equation}
\mu_{\pi}=\frac{m_\pi^2}{m_{u}+m_{d}}=1.5~GeV.
\end{equation}
 The amplitudes are calculated to be
\begin{eqnarray}
{\cal
A}^T(\overline{B}_{s}\rightarrow\pi^{+}\pi^{-})&=&\frac{G_{F}}{\sqrt{2}}
f_{B_s}f_\pi^2\pi\alpha_s(\mu)\frac{C_F}{N_C^2}C_1
\int_0^{\infty}dl_{+}\int_0^1dx\int_0^1dy\biggl\{\Phi_\pi(x)\Phi_\pi(y)\nonumber\\
& \times &
\biggl[\biggl(x\Phi_+^B(l_{+})+\xi\Phi_-^B(l_{+})\biggl)\frac{M_B^4}{D_sk_g^2}
+(\xi-y)\Phi_-^B(l_{+})\frac{M_B^4}{D_bk_g^2}\biggl]\nonumber\\
&+&\frac{\mu_\pi^2}{m_B^2}\phi_\pi(x)\phi_\pi(y)
\biggl[\biggl(x\Phi_+^B(l_{+})+y\Phi_-^B(l_{+})+3\xi\Phi_-^B(l_{+})
\biggr)\frac{M_B^4}{D_sk_g^2}+\biggl(\overline{x}\Phi_+^B(l_{+})\nonumber\\
&+&\overline{y}\Phi_-^B(l_{+})+3\xi\Phi_-^B(l_{+})
-2\frac{m_{b}}{m_{B}}\biggl(\Phi_+^B(l_{+})+\Phi_-^B(l_{+})\biggr)
\biggr)\biggr]\frac{M_B^4}{D_bk_g^2}\biggr\},
\label{am1}
\end{eqnarray}
\begin{eqnarray}
{\cal
A}^P(\overline{B}_{s}\rightarrow\pi^{+}\pi^{-})&=&\frac{G_{F}}{\sqrt{2}}
f_{B_s}f_\pi^2\pi\alpha_s(\mu)\frac{C_F}{N_C^2}
\int_0^{\infty}dl_{+}\int_0^1dx\int_0^1dy\biggl\{\Phi_\pi(x)\Phi_\pi(y)\nonumber\\
&\times&\biggl[\biggl(2C_4+\frac{C_{10}}{2}\biggl)
\biggl((x\Phi_+^B(l_{+})+\xi\Phi_-^B(l_{+}))
\frac{M_B^4}{D_sk_g^2}+(\xi-y)\Phi_-^B(l_{+})
\frac{M_B^4}{D_bk_g^2}\biggl)\nonumber\\
&+&\biggl(2C_6+\frac{C_{8}}{2}\biggl)
\biggl((\xi_B-x)\Phi_+^B(l_{+})+\xi\Phi_-^B(l_{+}))
\frac{M_B^4}{D_bk_g^2}\biggl)+y\Phi_-^B(l_{+})
\frac{M_B^4}{D_sk_g^2}\biggl]\nonumber\\
&+&\biggl(2C_4+2C_6+\frac{C_{8}}{2}+\frac{C_{10}}{2}\biggr)
\frac{\mu_\pi^2}{m_B^2}\phi_\pi(x)\phi_\pi(y)
\biggl[\biggl(\overline{x}\Phi_+^B(l_{+})\nonumber\\
&+&\overline{y}\Phi_-^B(l_{+})+3\xi\Phi_-^B(l_{+})
-2\frac{m_{b}}{m_{B}}\biggl(\Phi_+^B(l_{+})+\Phi_-^B(l_{+})\biggr)
\biggr)\frac{M_B^4}{D_bk_g^2}\nonumber\\
&+&\biggl(x\Phi_+^B(l_{+})+y\Phi_-^B(l_{+})+3\xi\Phi_-^B(l_{+})
\biggr)\frac{M_B^4}{D_s k_g^2}\biggr]\biggr\},
\label{am2}
\end{eqnarray}
where $\overline{x}=1-x$, $\xi_B=(M_B-m_b)/M_B$, and $\xi
=l_{+}/M_B$. $D_{b,s}$ and $k^2_g$ are the virtualities of b
quark, s quark and gluon propagators respectively.  $\Phi's $ are
the leading twist light-cone distribution amplitude(DA) of $\pi$
and B mesons. $\phi_\pi(x)$ is the twist-3 DA of $\pi$ meson.
These distribution amplitudes can be found in
Refs.\cite{Grozin,Genon,Braun,Huang} which describe long-distance
QCD dynamics of the matrix elements of quarks and mesons, which
are factorized out from the perturbative short-distance
interactions in the hard scatting kernels. For the distribution
functions of $B$ meson,  we use the model proposed
in~\cite{Grozin}
\begin{eqnarray}
\Phi_+^B(l_{+})&=&\sqrt{\frac{2}{\pi\lambda^{2}}}~\frac{l_+^2}{\lambda^{2}}~\exp\biggl
[-\frac{l_+^2}{2\lambda^{2}}\biggl],\\
\Phi_-^B(l_{+})&=&\sqrt{\frac{2}{\pi\lambda^{2}}}~\exp\biggl
[-\frac{l_+^2}{2\lambda^{2}}\biggl].
\end{eqnarray}
Now we can write the total decay amplitude
\begin{equation}
{\cal
A}(\overline{B}_{s}\rightarrow\pi^{+}\pi^{-})=V_{ub}V_{us}^*{\cal
A}^T-V_{tb}V_{ts}^*{\cal A}^P=V_{ub}V_{us}^*{\cal A}^T[1+
ze^{i(\gamma+\delta)}],
\end{equation}
where $z=|V_{tb}V_{ts}^*/V_{ub}V_{us}^*||{\cal A}^P/{\cal A}^T|$,
$\gamma=\arg[V_{tb}V_{ts}^*/V_{ub}V_{us}^*]$, $\delta$ is the
relative strong phase between penguin  and tree contribution
amplitudes, $z$ and $\delta$ can be calculate within QCD
factorization framework.

\begin{figure}
\begin{center}
\scalebox{0.8}{\epsfig{file=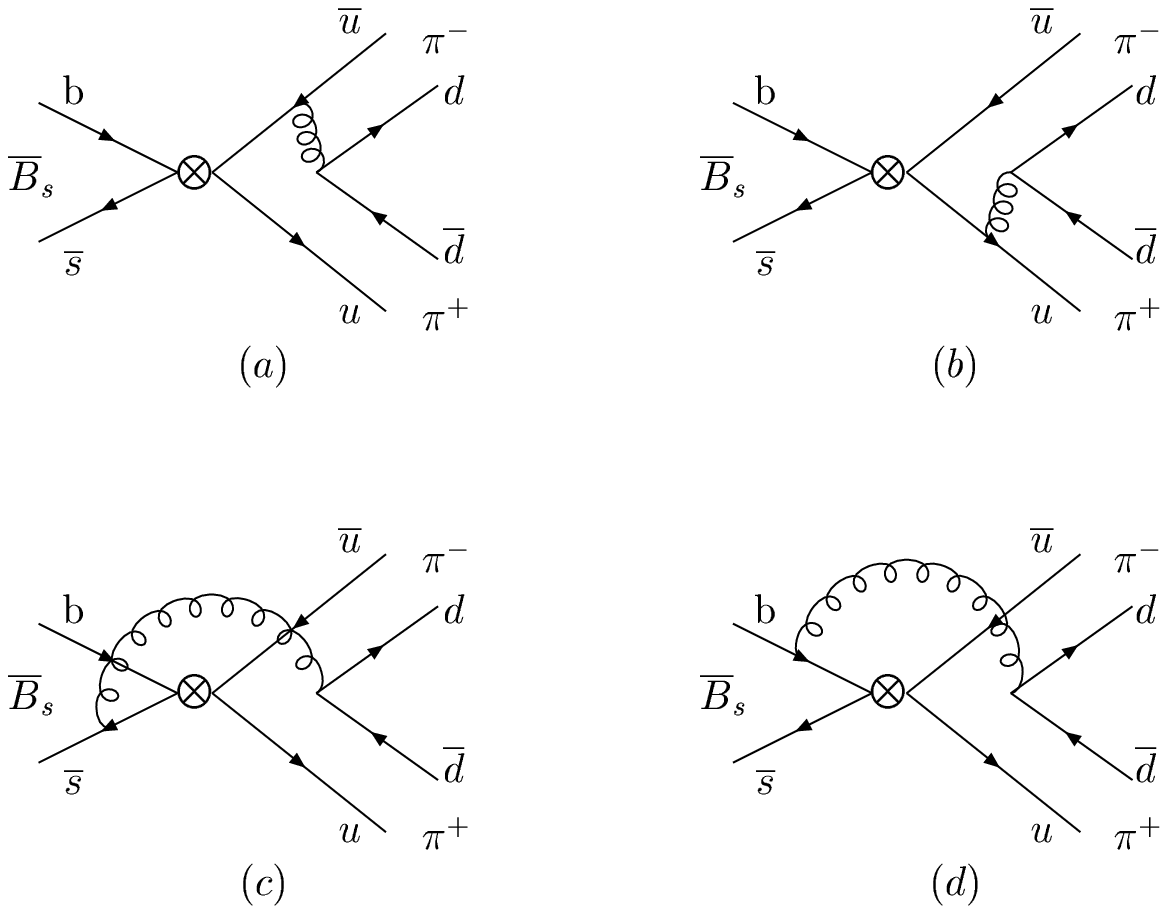}} \caption{\small The
annihilation diagrams for
$\overline{B}_{s}\rightarrow\pi^{+}\pi^{-}$ decay.}
\end{center}
\end{figure}

\section{Numerical results and Summary}

We list the parameters used in our numerical
calculation~\cite{pdg}
\begin{equation}
\begin{array}{llll}
M_{B_s}=5.37~GeV,& m_{b}=4.66~GeV,& \tau_{B_s^0}=1.461~ps,&
f_{B_s}=236~MeV,\\
  f_{\pi}=130~MeV,& \bar{\rho}=0.20,& \bar{\eta}=0.33.
\end{array}\label{ci}
\end{equation}
We set the scale in $\alpha_s (\mu)$ to be $M_{B_s}/2$ which is
about the averaged virtuality of the time-like gluon. In
Eq.\ref{am1}, \ref{am2} we meet endpoint divergences, which is the
known difficulty  to deal with the annihilation diagram within QCD
factorization framework. Instead of the widely used treatment
$\int^1_0 \frac{dy}{y}\to X_{A}=(1+\varrho_A
e^{i\varphi})\ln\frac{m_B}{\Lambda_h} $  in the
literature\cite{M.Beneke,Beneke,DU}, we  use an effective gluon
propagator~\cite{Cornwall} to avoid enhancements in the soft end
point region
\begin{equation}
\frac{1}{k^{2}}~\Rightarrow~\frac{1}{k^{2}+M_g^2(k^{2})},
\hspace{1cm}M_g^2(k^{2})=m_g^2 \biggl
[\frac{\ln(\frac{k^{2}+4m_g^2}{\Lambda^{2}})}
{\ln(\frac{4m_g^2}{\Lambda^{2}})}\biggl
]^{-\frac{12}{11}}.
\end{equation}
Typically $m_{g}=500\pm200$ MeV, $\Lambda=\Lambda_{QCD}$=250 MeV.
The use of this gluon propagator is supported by lattice result
~\cite{Williams}, and field theoretical
studies~\cite{thstudies,Alkofer} which have shown that the gluon
propagator is not divergent as fast as $\frac{1}{k^{2}}$.

For twist-3 DA $\phi_\pi(x)$, its asymptotic form is
$\phi_\pi(x)=1$ \cite{Braun} which used in \cite{bn,DU}.  To
further suppress endpoint contributions, we will use   the recent
model by Huang and Wu\cite{Huang}
\begin{equation}
\phi_{\pi}(x)=\frac{A_p \beta^2}{2\pi^2}\left[ 1+B_{p}C^{1/2}_2
(1-2x)+C_p C^{1/2}_4 (1-2x)\right] \exp\left[-\frac{m^2}{8\beta^2
x(1-x)}\right],
\end{equation}
where $C^{1/2}_2 (1-2x)$ and $C^{1/2}_4 (1-2x)$  are Gegenbauer
polynomials and other parameters could be found in \cite{Huang}.

Using these inputs, we get the CP averaged branching ratio of the
decay
\begin{equation}
Br(\bar{B}_{s}\to \pi^{+}\pi^{-})=(1.24\pm0.28)\times10^{-7}.
\end{equation}
 The available upper limit of the decay
at 90\% confidence level~\cite{pdg} is
\begin{equation}
Br(\overline{B}_{s}\rightarrow\pi^{+}\pi^{-})<1.7\times10^{-4}.
\end{equation}
Obviously, our result is far below this upper limit. However, our
result is  larger than these QCD factorization result
$Br(\bar{B}_{s}\to \pi^{+}\pi^{-})\simeq2\times 10^{-8}$
\cite{bn,DU} by using the treatment $\int^1_0 \frac{dy}{y}\to
X_{A}=(1+\varrho_A e^{i\varphi})\ln\frac{m_B}{\Lambda_h} $. We
also note that our result may consistent with the one of
Ref.\cite{bn} $Br(\bar{B}_{s}\to
\pi^{+}\pi^{-})=(0.024^{+0.003+0.025+0.163}_{-0.003-0.012-0.021})\times
10^{-6}$ if the huge uncertainties   are considered.   In a recent
study in the framework of PQCD factorization\cite{L.Ying}, the
authors found $Br(\bar{B}_{s}\to \pi^{+}\pi^{-})=(4.2\pm
0.6)\times 10^{-7}$ where the end point divergence is regulated by
$k^2_{\perp}$.

\begin{figure}
%[htbp]
\begin{tabular}{cc}
\scalebox{0.7}{\epsfig{file=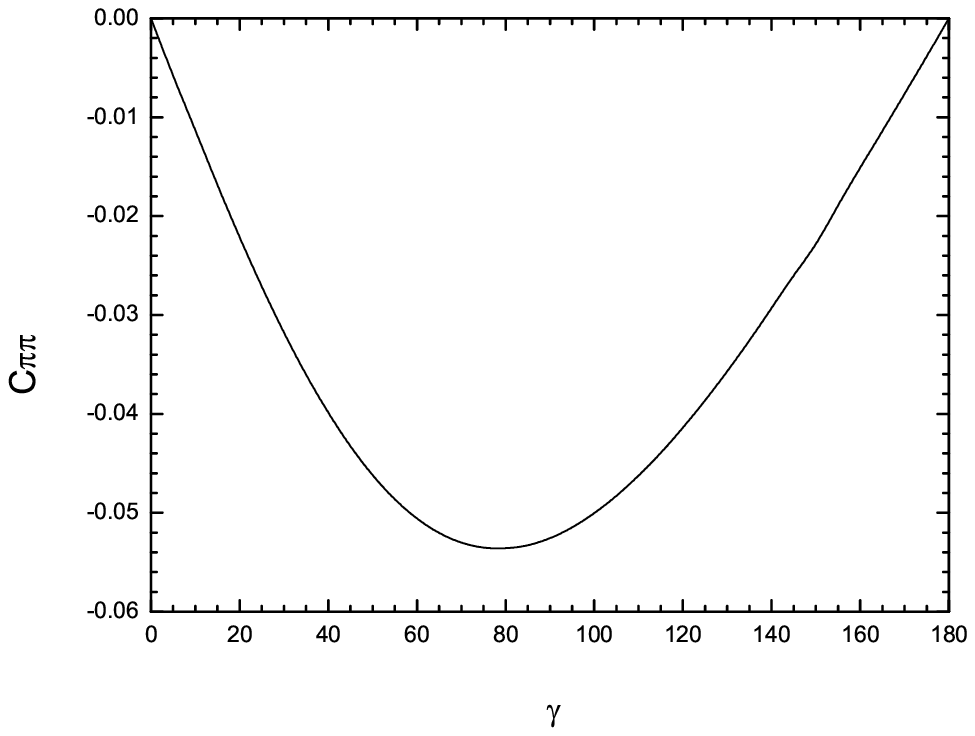}} & \scalebox{0.7}{\epsfig{file=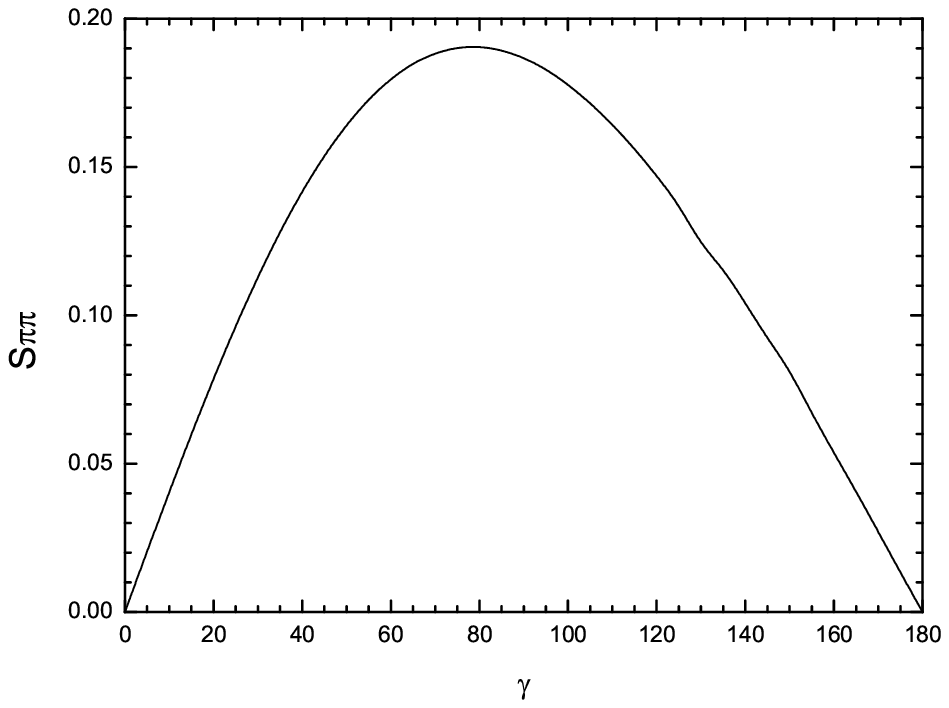}}  \\
(a)&(b)
%    \\[4mm]
\end{tabular}
\caption{\small The direct
 CP violation parameter  $C_{\pi\pi}$ and the
mixing-induced CP violation parameter $S_{\pi\pi}$ of
$\overline{B}_{s}\rightarrow\pi^{+}\pi^{-}$ decay as a function of
weak phase $\gamma$(in degree).}
%\end{center}
\end{figure}

The absolute ratio between the amplitude of penguin  and the tree
is $z=9.8$,   and the strong phase is $\delta=164^{\circ}$.  So,
we can see that almost all the contribution comes from penguin.
Our results for z and $\delta$ agree with the PQCD
results\cite{L.Ying}.

Now it is time to discuss CP asymmetries of $\bar{B}_s (B_s)\to
\pi^+ \pi^- $. The time-dependent asymmetries are given
by\cite{gronau}
\begin{equation}
A_{CP}(t)\equiv \frac{\Gamma(\bar{B}_s (t)\to \pi^+ \pi^-)-
\Gamma(B_s (t)\to \pi^+ \pi^-)}{ \Gamma(\bar{B}_s (t)\to \pi^+
\pi^-)- \Gamma(B_s (t)\to \pi^+ \pi^-)}=C_{\pi\pi}\cos(\triangle
mt)+ S_{\pi\pi}\sin(\triangle mt),
\end{equation}
where $\triangle m$ is the mass difference of the two mass
eigenstates of $B_{s}$ meson. $C_{\pi\pi}$ and $S_{\pi\pi}$ are
parameters describing the direct CP violation and the
mixing-induced CP violation, respectively.

Finally, our results for direct and mixing-induced CP violations
in the decay are presented as  functions of weak phase $\gamma$ in
Fig.2.a, b respectively.  For $\gamma=60^{\circ}\pm
14^{\circ}$\cite{pdg}, the direct CP violation parameter
$C_{\pi\pi}$  is about -0.05, the mixing-induced CP violation
parameter $ S_{\pi\pi}$ of the decay is as large as 0.18.

In summary, we have calculated the CP averaged branching ratio and
CP asymmetries of the decay
$\overline{B}_{s}\rightarrow\pi^{+}\pi^{-}$ within the framework
of QCD factorization. We have obtained  that the CP averaged
branching ratio of this decay mode is of the order of $10^{-7}$.
The CP violations are estimated to be $C_{\pi\pi}=-0.05$,
$S_{\pi\pi}=0.18$.  Compared with former studies in the same
framework, we have included both the two distribution functions
$\Phi^{B}_{+}$ and $\Phi^{B}_{-}$ of $B_s$ meson. We also have
used Cornwall prescription\cite{Cornwall} for the gluon propagator
with a dynamical mass to avoid enhancements in soft endpoint
region.  It is noted that recent studies\cite{thstudies,phstudies}
have given support for Cornwall prescription, which might have
many phenomenological applications in B decays. Once future
measurements at LHCb in agreement with our predictions, it would
indicate that Cornwall prescription could be used in QCDF to
improve it's treatment of endpoint divergences in hard-spectator
scattering and annihilation topologies to enhance its power for
analyzing charmless B nonleptonic decays.

\section*{Acknowledgments}

This work is supported  by NFSC under contract No.10305003, Henan
Provincial Foundation for Prominent Young Scientists under
contract No.0312001700 and in part by the project sponsored by SRF
for ROCS, SEM.

\end{document}